# $^{55}$Mn NMR in Mn$_{12}$ acetate: Hyperfine interaction and magnetic relaxation of cluster

T. Kubo*

*Physics Department, Faculty of Education, Nara University of Education, Nara 630-8528, Japan*

T. Goto and T. Koshiba

*Graduate School of Human and Environmental Studies, Kyoto 606-8501, Kyoto University, Japan*

K. Takeda

*Department of Chemistry, Graduate School of Science, Sapporo 060-0810, Hokkaido University, japan*

K. Awaga

*Department of Chemistry, Graduate School of Science, Nagoya University, Nagoya 464-8602, Japan*

**Abstract**

The $^{55}$Mn NMR in oriented powder crystal of Mn$_{12}$Ac has been investigated at 1.4−2.0 K in zero field and with external fields along the *c*-axis. Three kinds of $^{55}$Mn NMR, with central frequencies at 230.2±0.1, 279.4±0.1 and 364.±0.1 MHz, composed of five-fold quadrupole-split lines for *I*=5/2 nuclei have been interpreted to arise from Mn$^{4+}$ ion, and two crystallographically-inequivalent Mn$^{3+}$ ions, respectively. It is found that the isotropic hyperfine field in the Mn$^{4+}$ ion with 3$d^3$ configuration indicates a large amount of reduction (26%) as compared with the theoretical evaluation. In the analysis for the hyperfine field of Mn$^{3+}$ ions with 3$d^4$ configuration, we have taken into account of the anisotropic dipolar contribution in addition to the Fermi-contact term in order to explain two kinds of $^{55}$Mn NMR frequencies in Mn$^{3+}$ ions in inequivalent sites. Using the reduction factor for the magnetic moment determined by polarized neutron diffraction experiment, we obtained the reduction factor of 8% from the calculated values for the dipolar term so as to fit the observed frequency. We suppose that such an appreciable amount of the reduction factor is due to covalence and strong exchange interaction among manganese ions via the oxygen ions. By using the hyperfine coupling constants of twelve manganese ions in Mn$_{12}$Ac, the total hyperfine interaction of the ferrimagnetic ground state of *S*=10 has been determined to amount to 0.3 cm$^{-1}$ in magnitude at most, the magnitude of which corresponds to the nuclear hyperfine field $h_e$=0.32 kG seen by Mn$_{12}$ cluster spin.

The relaxation of the cluster magnetization was investigated after reversal of the external field by observing the recovery of the $^{55}$Mn spin-echo intensity in the fields of 0.20−1.90 T along the c-axis at 2.0 K.  It was found that the magnetization of the cluster exhibits the √*t*-recovery in the short time regime. The relaxation time decreases with increasing external field following significant dips at every 0.45 T. This is interpreted to be due to the effects of thermally-assisted quantum tunneling between the spin states at magnetic level crossings.

## I. INTRODUCTION

Recently there has been a great interest in the compounds of nano-scale magnets with high value of the spin involving a large number of magnetic ions. [1-6] This is based on the view of studying the quantum physics of meso-scopic system, which lies in the intermediate scale between the classical macroscopic and quantum microscopic scales. Among various molecular clusters, $Mn_{12}O_{12}(CH_3COO)_{16}(H_2O)_4$ (abbreviated as $Mn_{12}Ac$) [7] has been most extensively studied theoretically [8-16] and also experimentally so far using magnetization, [17-26] ac-susceptibility, [27,28] heat capacity, [29,30] neutron-scattering, [31-34] ESR, [35,36] and proton NMR and μSR. [37,38] The most prominent feature is the characteristic magnetization relaxation and the step-wise hysteresis curve associated with the phonon-assisted resonant quantum tunneling. [19,20,39,40]

In $Mn_{12}Ac$, each of the magnetic clusters, which constitute a tetragonal symmetry ($S_4$), is constructed from four $Mn^{4+}$ ($S=3/2$) ions (the site denoted as Mn1) in a central tetrahedron and surrounding eight $Mn^{3+}$ ($S=2$) ions with two crystallographically inequivalent sites (Mn2 and Mn3) which are located alternately in the outer ring. [7] Figures 1(a) and (b) show respectively the molecular structure viewed from the crystallographical *c*-axis, and the two inequivalent $Mn^{3+}$ sites surrounded octahedrally by neighboring oxygens together with their local tetragonal axes δ.

Magnetic ground-state of $Mn_{12}Ac$ cluster is determined by combined effect of four paths of intramolecular exchange interactions, whose values are estimated as follows in Ref. 18: the dominant interaction between Mn1-Mn2 is antiferromagnetic (−150 cm$^{-1}$) and the interactions between the neighboring two Mn1 ions and between Mn1-Mn3 are both also antiferromagnetic (−60 cm$^{-1}$). The interaction between Mn2-Mn3 is ferromagnetic or antiferromagnetic (less than 40 cm$^{-1}$). Because of the single ion anisotropy originating in $Mn^{3+}$ ions due to Jahn-Teller effect, the cluster possesses the total Ising-like anisotropy along the crystallographic *c*-axis. The following model is established experimentally: four $Mn^{4+}$ spins in the inner tetrahedron are parallel to each other ($S=3/2\times4=6$) and eight $Mn^{3+}$ spins in the outer ring are also parallel ($S=2\times8=16$). Then total spins of two groups couple antiparallelly to form $S=10$ for the cluster. [17] The most of magnetic properties have been well understood in terms of the doubly-degenerate and discrete energy levels of the cluster with a high-spin of $S=10$, which are specified by wave function $|S,m\rangle$ in terms of the following Hamiltonian, $H_0=-DS_z^2$ where the constant $D$ is the effective Ising-like anisotropy of the order of 0.52–0.60 K [33,35] along the *c*-axis. In addition to this term, it has been recognized that there are the quartic terms denoted $H'$, one of which plays an important role for the tunneling, though the magnitude is much smaller as compared with the main term. Then, in the presence of an external field $H$ applied along the $c$-axis, the spin Hamiltonian describing the cluster molecule is given as

$$H = H_0 - g_z\mu_B S_z H + H = -DS_z^2 - g_z\mu_B S_z H - ES_z^4 + B(S_+^4 + S_-^4) \qquad (1)$$

where $g_z$=1.93,[35] $E$= (1.1–1.3)×10$^{-3}$ K [32,33] and $B$=±(2.1–7.2)×10$^{-5}$ K.[32,35]

On the other hand there has been proposed another spin model,[18] which describes that the Mn$_{12}$ molecular spin ($S$=10) is composed of four dimers with $s$=1/2 (an antiferromagnetically coupled pair of Mn1-Mn2) and four $S$=2 (Mn$^{3+}$ of Mn3). This model has been used for theoretical interpretation [41,42] of magnetic properties according to simple description of the exchange interactions.

Considering that the spin structure of the high-spin is basically constituted of four Mn$^{4+}$ ions and of eight Mn$^{3+}$ ions, it seems also important to investigate, on the microscopic point of view, the ground-state of each manganese ion, in order to understand the nature of the ground-state of the cluster. The most prominent experimental procedure for this purpose is to examine $^{55}$Mn NMR in Mn$_{12}$Ac. The NMR information on behavior of manganese ions will be also important to interpret magnetic relaxation with quantum tunneling at low temperature. On such a view point, we have been studying the $^{55}$Mn NMR in Mn$_{12}$Ac. We have already succeeded in observing three kinds of $^{55}$Mn NMR for powder crystals of Mn$_{12}$Ac below 2.4 K in zero external field. In the preliminary paper,[43] we have assigned the lowest $^{55}$Mn NMR spectra as due to Mn$^{4+}$ ions while the higher two ones to crystallographically-inequivalent Mn$^{3+}$ ions, and reported the result for nuclear relaxation time $T_2$. Subsequently we investigated the relaxation of magnetization of Mn$_{12}$ cluster by monitoring spin-echo intensity of Mn$^{4+}$ ions in the external field applied along the c-axis.[44]

In the present work we have performed $^{55}$Mn NMR measurements in more detail on an oriented-powder sample of Mn$_{12}$Ac in zero field and with applied field along the $c$-axis. By considering anisotropic terms of the hyperfine fields due to the dipolar interaction in addition to Fermi-contract term we have analyzed the hyperfine fields both of Mn$^{4+}$ ions and of two inequivalent Mn$^{3+}$ ions. We have explained results in accurate assignment of three kinds of $^{55}$Mn NMR spectra to relevant manganese ions located at Mn1, Mn2 or Mn3 sites, respectively. In order to investigate the behavior of magnetization recovery and quantum tunneling in Mn$_{12}$Ac we have also measured the relaxation time of cluster magnetization as a function of the external field, using the spin-echo signal of Mn$^{4+}$ NMR in the field of 0.20–1.90 T along the $c$-axis at 2.0 K.

## II. EXPERIMENTAL RESULTS
### A. $^{55}$Mn NMR

The experiment has been performed by the conventional spin-echo method using pulsed NMR apparatus. The oriented powder sample of Mn$_{12}$Ac contained in a capsule (6 mm$\phi$ and 20 mm in length) is inserted in a resonant cavity of cylindrical shape.

Figure 2 shows the experimental results of $^{55}$Mn NMR spectra obtained at $T$=1.4 K in the frequency range between 150 and 450 MHz in zero-external field. Each of three kinds of $^{55}$Mn NMR is composed of five-fold quadrupole-split lines for $I$=5/2 nucleus. The width of the spectra around 230 MHz is nearly four times narrower than those of others. As we explain in detail in the next section, we assign these three kinds of NMR spectra to manganese nuclei associated with Mn$^{4+}$ ion and two inequavalent Mn$^{3+}$ ions by taking into account of individual resonance profiles such as a central frequency, quadrupole-splitting and field-dependence, and by considering the anisotropy of hyperfine field. In this section, we mention briefly the results for assignment of three NMR spectra.

The spectra around 230 MHz in Fig. 2 correspond to $^{55}$Mn NMR in Mn$^{4+}$ ions at Mn1 site. The central frequency is 230.2±0.1 MHz with a quadrupole-splitting $\nu_Q$ =0.72±0.05 MHz which can be noticed in the inset. This quite small quadrupole-splitting is attributed to the deformation of the crystal field from cubic symmetry for an orbital singlet ion. The resonance frequency corresponds to the hyperfine field $|H_N|$=218.4±0.1 kG. On the other hand, in the case of an orbital doublet ion, the NMR spectra shows generally relatively large quadrupole-splitting coming from the aspherical distribution of outer electrons. So the two broad spectra are referred to the orbital doublet ion, i.e. the Mn$^{3+}$ ion. As described later in Sec. III, the magnitude of the hyperfine field for the Mn$^{3+}$ ion increases, owing to anisotropic dipolar contribution, in accordance with the increase in the angle $\theta$ between the electron spin axis and the local tetragonal axis. On the other hand, the quadrupole splitting decreases as the angle $\theta$ increases. Since the angle $\theta$ is larger in the Mn3 site than the one for the Mn2 site, the hyperfine field and hence $^{55}$Mn NMR frequency are larger for the Mn$^{3+}$ ion at the Mn3 site while the quadrupole splitting is smaller. Then we may assign the broad spectra centered at 279.4±0.1 MHz ($|H_N|$=264.8±0.1 kG) with $\nu_Q$=4.3±0.1 MHz to the Mn$^{3+}$ ion at the Mn2 site, and the spectra 364.4±0.1 MHz ($|H_N|$=345.4±0.1 kG) with $\nu_Q$ = 2.9±0.1 MHz to the Mn$^{3+}$ ion at the Mn3 site, respectively.

Figure 3 shows the external magnetic field dependence of each central line of the above-ascertained $^{55}$Mn NMR, which has been measured at $T$= 1.4 K in the field $H$ applied along the $c$-axis. We obtained $\gamma$-values which coincide with that of free $^{55}$Mn nucleus ($^{55}\gamma/2\pi$= 1.055 MHz/kG) within the experimental error. When the specimen is cooled down without an external field (denoted as ZFC), we observe two components for every central frequency, i.e. increasing and decreasing branches versus $H$. The typical NMR spectra of $^{55}$Mn NMR in Mn$^{4+}$ ions are shown in the inset to Fig. 3, where the narrow line (dashed line) in the zero field splits into two components $\nu_u$ and $\nu_d$ at $H$=1.0 T after ZFC. However, when the specimen is cooled down below the blocking temperature $T_B$= 3 K with $H$ parallel to the $c$-axis (we denote this case by FC), we observe only one branch expressed by the solid line.

We interpret the present $^{55}$Mn NMR frequency-field diagram by considering the magnetic structure of the cluster. Due to the uniaxial anisotropy along the $c$-axis, the magnetic moment of $S$=10 stays along the $c$-axis even if there is no external field. Because of weak dipolar interaction (~0.05 K) [40] among neighboring clusters, each molecular cluster, however, remains superparamagnetic. Therefore, when it is cooled down with ZFC the molecular magnets have two possibilities in direction, parallel or antiparallel to the $c$-axis in each crystal. That is the reason why we observed two branches shown by solid and dotted lines in Fig. 3 in the field dependence of $^{55}$Mn NMR in the case of ZFC. On the other hand when the whole molecular crystals are cooled down in FC, they have a magnetic moment orienting to the unique direction along $H$, shown in Fig. 4(a). Considering that $^{55}$Mn NMR frequency is proportional to the total field acting on $^{55}$Mn nucleus, the three solid lines in Fig. 3 mean that effective fields acting on $^{55}$Mn nuclei in two different Mn$^{3+}$ ions decrease with the external field, while those acting on $^{55}$Mn nuclei in Mn$^{4+}$ ions increase. Considering both hyperfine fields to be oppositely directed to their local moments for Mn$^{3+}$ and Mn$^{4+}$ ions, the fact that we observed only the solid lines as shown in Fig. 3 in the case of FC is consistent with the fact that the magnetic moment of the Mn$^{3+}$ ions are parallel to the total moment of the cluster while those of the Mn$^{4+}$ ions are antiparallel, [18] as shown in Fig. 4(a) and 4(c). The situation is explained in detail in Sec. A and B.

## B. Magnetic relaxation of Mn$_{12}$Ac cluster

The magnetic relaxation of the cluster was measured by cooling the oriented powder sample down to $T$=2.0 K from $T$=4.2 K under the applied field of $H$=1.2 T along the $c$-axis. After this FC process, we can observe $^{55}$Mn NMR in Mn$^{4+}$ ions at $\nu$=230.4+($\gamma_{Mn1}/2\pi$)$H$= 243.0 MHz in the field of $H$=1.2 T, while we can not observe any signal at $\nu$=230.4−($\gamma_{Mn1}/2\pi$)$H$= 217.8 MHz. The situation is upset after reversing the field, i.e. we observe $^{55}$Mn NMR in Mn$^{4+}$ ions only at $\nu$=217.8 MHz while we can not observe any signal at $\nu$ = 243.0 MHz. The reversal of the external field can be reached within a few seconds. Accordingly, by reversing the external field after FC, we can measure the relaxation of the cluster magnetization in the field by probing the growth of spin-echo amplitude of $^{55}$Mn NMR in Mn$^{4+}$ ions at $\nu$= 243.0 MHz. [44]

Figure 5 shows the recovery of the spin-echo intensity of $\nu_n$ branch of $^{55}$Mn NMR in Mn$^{4+}$ ion after the reversal of the external field, where the time-dependent growth of the intensity of $\nu_n$ line corresponds to magnetic relaxation of cluster moment along the applied field. The typical example of relaxation curve is depicted in Fig. 5, in which the growth of spin-echo intensity was shown for two values of $H$ =1.2 T and 1.4 T at $T$= 2.0 K. From the recovery of the signal intensity we can recognize the time-dependence of $\sqrt{t}$-type until about 20 min for $H$=1.2 T and 13 min for $H$=1.4 T. After that duration the single exponential recovery follows. From the slope of the $\sqrt{t}$-recovery we can get the relaxation time $\tau$ measured in short-time regime as a function of $H$. In order to see the effect of the quantum tunneling, we performed measurements by changing fields by steps of 0.025 T between 0.200 T and 1.900 T. The inset demonstrates that $\tau$ shows significant dips of more than one order of magnitude at every 0.45 T step with a tendency to decrease with increasing the fields. Such a feature is obvious above 2.0 K, while below about 1.8 K it not so clear.

### III. HYPERFINE INTERACTION AND $^{55}$Mn NMR
#### A. Hyperfine interaction and hyperfine field

First we explain a general treatment for evaluation of the hyperfine field acting on the nucleus of a magnetic ion in magnetic material. The nucleus with a nuclear magnetic moment $\gamma\hbar\mathbf{I}$ and a quadrupole-moment $eQ$ in a transition metal ion with an unpaired spin $\hbar\mathbf{S}$ and an orbital moment $\hbar\mathbf{l}$ is subject to the hyperfine interaction $H_{hyp}$ which is composed of interaction with its own electrons via a magnetic hyperfine interaction $H_n$ and an electrostatic quadrupole-interaction $H_Q$, namely $H_{hyp}= H_n+ H_Q$.[46,47]

In general, magnetic hyperfine interaction of nucleus consists of Fermi-contact $H_F$, dipolar $H_d$, and orbital interactions $H_l$,

$$H_n=H_F+H_d+H_l \quad (2)$$

which are expressed, using usual notations,[48] respectively, as

$$H_F = -\frac{16\pi}{3}\gamma\hbar\mu_B \mathbf{I}\sum_k \delta(\mathbf{r}_k)\mathbf{s}_k, \quad (3)$$

$$H_d = 2\gamma\hbar\mu_B \sum_k \{-\frac{\mathbf{s}_k \cdot \mathbf{I}}{r_k^3} + \frac{3(\mathbf{r}_k \cdot \mathbf{I})(\mathbf{r}_k \cdot \mathbf{s}_k)}{r_k^5}\} \quad (4)$$

and

Here the summation k in the first Eq. (3) is taken over all inner-core *s*-electrons and the summations in the second Eq. (4) and third Eq. (5) are taken over all the magnetic 3*d*-

$$H_l = 2\gamma\hbar\mu_B \sum_k \frac{\mathbf{l}_k \cdot \mathbf{I}}{r_k^3}. \tag{5}$$

electrons within the ion to which the referring nucleus belongs. The dipolar interactions with the other surrounding magnetic ions are negligibly small as compared with the above Eq. (4).

The electrostatic quadrupole-interaction $H_Q$ is given by the following equation,[47,48]

$$H_Q = \frac{eQ}{2I(I-1)} \sum_k \left\{ \frac{\mathbf{I}^2}{r_k^3} - \frac{3(\mathbf{r}_k \cdot \mathbf{I})^2}{r_k^5} \right\}. \tag{6}$$

In the case where the magnetic ion is surrounded by anions with octahedral configuration, it is convenient to define the (*X, Y, Z*) co-ordinate system referred to the three principal axes of the electric field gradient with the *Z*-axis of the highest rotation symmetry. The effective magnetic hyperfine Hamiltonian is obtained by averaging $H_n$ with respect to the electronic ground-state wave function $\psi_g$, and can be expressed in terms of the hyperfine field $\mathbf{H}_N$ acting on the nucleus as follows;[46,49]

$$\langle\psi_g|H_n|\psi_g\rangle = \langle\psi_g|H_F|\psi_g\rangle + \langle\psi_g|H_d|\psi_g\rangle + \langle\psi_g|H_l|\psi_g\rangle = -\gamma\hbar\mathbf{I}\cdot\mathbf{H}_N = -\gamma\hbar\mathbf{I}\cdot(\mathbf{H}_F + \mathbf{H}_d + \mathbf{H}_l) \tag{7}$$

in which $\mathbf{H}_F$, $\mathbf{H}_d$ and $\mathbf{H}_l$ are Fermi-contact, dipolar and orbital hyperfine fields resulting from the corresponding terms of the magnetic hyperfine interactions (3)–(5), respectively.

On the other hand the effective nuclear-quadrupole Hamiltoninan is obtained by the expectation value of $H_Q$ with respect to $\psi_g$, as follows,[50]

$$\langle\psi_g|H_Q|\psi_g\rangle = \frac{e^2 Q q_{ZZ}}{4I(I-1)}\{3I_Z^2 - I(I+1) + \eta(I_X^2 - I_Y^2)\} \tag{8}$$

where $eq_{ZZ}$ is the maximum principal value of the electric field gradient tensor $V_{ZZ}$ along the *Z*-axis,[46] and $\eta$ is an asymmetric parameter defined by $(V_{XX}-V_{YY})/V_{ZZ}$. In some cases, the effect of the surrounding anions themselves becomes appreciable.
Accordingly the effective total hyperfine Hamiltonian is expressed to be

$$\langle\psi_g|H_{hyp}|\psi_g\rangle = \langle\psi_g|H_n|\psi_g\rangle + \langle\psi_g|H_Q|\psi_g\rangle. \tag{9}$$

The Fermi-contact field, which arises from the difference between the positive and negative spin densities at the nucleus due to core-polarization of inner *s*-electrons caused by 3*d*-electron spins, is dominant in magnitude ($\approx 10^2$ kG) and opposite in direction to the atomic magnetic moment.[47] The dipolar hyperfine field exists in the magnetic ion whose ground-state is degenerate in a cubic field, while it does not exist in a singlet ion.[46,47] Concerning the orbital hyperfine field it is normally zero for the case of an orbital singlet state or of a doublet state, since the orbital angular momentum is quenched. However it survives in a little amount when the ground orbital state is admixed with excited states through the spin-orbit interaction and/or low-symmetry crystal-field. In this case its magnitude depends only on the *g*-shift. Namely the orbital hyperfine field is described as [47]

$$H_l = -2\mu_B \langle r^{-3} \rangle \Delta g. \qquad (10)$$

The ground state of the $Mn^{4+}$ ion ($3d^3, {}^4F$) is orbital singlet [46] in a cubic crystal field $V_c$. The dipolar hyperfine field of the $Mn^{4+}$ ion vanishes because the orbital wave-function and hence also the distribution of spin magnetization have cubic symmetry. It also follows that quadrupole splitting vanishes for an orbital singlet $Mn^{4+}$ ion in $V_c$ since $V_{XX}=V_{YY}=V_{ZZ}=0$ resulting from Laplace equation $V_{XX}+V_{YY}+V_{ZZ}=0$ in the case of cubic symmetry of the surroundings.[47]

In general the *g*-factor for $Mn^{4+}$ has been observed to be $\approx 1.99$ and $|\Delta g|$ appears normally less than 0.01,[46] so the orbital hyperfine field is evaluated to be less than 10 kG from the relation $H_l = -2\mu_B \langle r^{-3} \rangle \Delta g = -125 \langle r^{-3} \rangle_{a.u.} \Delta g$ kG, using the value $\langle r^{-3} \rangle_{a.u.}=5.36$ expressed in atomic units for free $Mn^{4+}$ ions.[46] Then the orbital hyperfine field can be reasonably neglected compared with $H_F$. Therefore the hyperfine field of the $Mn^{4+}$ ion is mainly composed of the Fermi-contact term, which is isotropic and oppositely directed to the magnetic moment of the ion, and parallel to the electron spin **S**. Usually the hyperfine interaction is effectively denoted as $A$ **S**·**I**, which corresponds to the Zeeman interaction of nuclear magnetic moment $\gamma\hbar$**I** subject to the Fermi-contact field **H**$_F$. Then the hyperfine field is simply expressed to be,

$$\mathbf{H}_N = \mathbf{H}_F = -(A/\gamma\hbar)\mathbf{S}, \qquad (11)$$

where $A$ is an isotropic hyperfine coupling constant. So, the $^{55}Mn$ NMR frequency of the $Mn^{4+}$ ions given by $\nu = \gamma H_N/2\pi\hbar$ is proportional to $|\mathbf{S}|$ and its $^{55}Mn$ NMR spectra are to have width narrower than those of the $Mn^{3+}$ ions. The relation between **H**$_N$ and **S** together with magnetic moment **m** is shown in Fig. 4(c).

As for the hyperfine field for the $Mn^{3+}$ ion ($3d^4, {}^5D$) the situation is somewhat complicated owing to the orbital degeneracy in $V_c$.[46] The degeneracy of an orbital doublet $E_g$ is removed to two orbital states denoted as $A_{1g}$ and $B_{1g}$ by the deformation of the crystal field to tetragonal one caused mainly by Jahn-Teller effect. In addition there arises some crystallographic influence of low-symmetric carboxylate ligands [7] in the molecular cluster. The ground state $B_{1g}$ is specified by wave-function $|X^2-Y^2\rangle$ of a hole plus a half-filled shell when the tetragonal distortion is of a prolate type. In case an orthorhombic distortion is also superposed the double-fold $E_g$ states are not only split but also admixed, so that the orbital ground state is of the form [46]

$$\psi_g = \cos\phi \, |X^2-Y^2\rangle + \sin\phi \, |3Z^2-r^2\rangle. \tag{12}$$

When this type of the ground-state wave function is applied to Eq. (7) we obtain an appreciable amount of the dipolar field at the $^{55}$Mn nucleus.

As given in Appendix A the dipolar hyperfine field for the $Mn^{3+}$ ion is calculated from Eq. (4) by using the ground-state wave function (12), and the result is $\mathbf{H}_d = \mathbf{D}\,\hat{\mathbf{e}}_m$, where $\hat{\mathbf{e}}_m$ is a unit vector of magnetization $\mathbf{m}$ of $Mn^{3+}$ ion and $\mathbf{D}$ is the dipolar hyperfine field coupling tensor. Therefore when the singlet ground-state of the $Mn^{3+}$ ion is specified by the wave function (12) the total hyperfine field is expressed as follows, [49]

$$\mathbf{H}_N = \mathbf{H}_F + \mathbf{H}_d = -|H_F|\hat{\mathbf{e}}_m + \mathbf{D}\,\hat{\mathbf{e}}_m \tag{13}$$

It is noted that the orbital hyperfine field for the $Mn^{3+}$ ion can be neglected here since $H_l$ is evaluated to be $-2\mu_B\langle r^{-3}\rangle\Delta g = -125\langle r^{-3}\rangle_{a.u.}\Delta g = -6$ kG, using $\langle r^{-3}\rangle_{a.u.} = 4.79$ a.u. [47] and $\Delta g \approx 0.01$.[51] Then the total hyperfine field $\mathbf{H}_N$ expressed by the Eq. (17) is approximately in opposite direction to $\mathbf{m}$ and directs against the local tetragonal axis $\delta$ with the following angle $\alpha$,[52]

$$\tan\alpha = \frac{H_F - \tfrac{1}{2}H'_d}{H_F + H'_d}\tan\theta. \tag{14}$$

The relation between $\alpha$ and $\theta$ together with $\mathbf{m}$, $\mathbf{S}$ and $\mathbf{H}_N$ is shown in Fig. 4(c).

### B. $^{55}$Mn NMR frequency

In the analysis of $^{55}$Mn NMR frequency expected from the hyperfine interaction (9), it is convenient to use another reference frame $xyz$ where $z$-axis is parallel to the direction of the magnetic moment of the ion, i.e. $c$-axis, and $x$-axis is in the plane containing $z$- and $Z$-axis as shown in Fig. 4(b), where $\theta$ is an angle between the magnetic moment and the $Z$-axis. We assume $\eta=0$, then the effective Hamiltonian (9) is transformed to be as follows,

$$\langle\psi_g|H_{hyp}|\psi_g\rangle = -\gamma\hbar\mathbf{I}\cdot(\mathbf{H}_F + \mathbf{H}_d) + \{e^2q_{ZZ}Q/4I(2I-1)\}[3I_z^2 - I(I+1)]$$

$$= -\gamma\hbar[H_F\cdot I_z - (1/2)(3\cos^2\theta - 1)H_d \cdot I_z - (3/4)\sin2\theta\, H_d \cdot I_x]$$
$$+ e^2q_{ZZ}Q/\{4I(2I-1)\}[(1/2)(3\cos^2\theta-1)\{3I_z^2 - I(I+1)\}$$
$$+ (3/2)\sin2\theta(I_z I_x + I_x I_z) + (3/2)\sin^2\theta(I_x^2 - I_y^2)], \quad (15)$$

where the off-diagonal terms of Zeeman- and quadrupole-terms are perturbations to their respective main terms.[49]

Taking into account the main term in (15), the NMR frequency $\nu_m$ corresponding to the transition m−1↔m is given by

$$\nu_m = |E_{m-1} - E_m|/h = \nu_F - (1/2)(3\cos^2\theta - 1)\{\nu_d - (m-1/2)\nu_Q\} + \text{higher-order terms}, \quad (16)$$

where $\nu_F$ and $\nu_d$ are obtained from the isotropic part of the hyperfine field $H_F$ and the anisotropic one $H_d$, respectively, by the relation $\nu = \gamma|H|/2\pi\hbar$ between frequency and field, and $\nu_Q$ is the quadrupole frequency $3e^2q_{ZZ}Q/\{2I(2I-1)h\}$. The higher-order terms, which are related to the off-diagonal term in (15), are evaluated to be less than a few per cent compared with main terms.

In the case of the $Mn^{4+}$ ion the dipolar hyperfine field is zero, then $\nu_m$ is simply to be

$$\nu_m = \nu_F + (1/2)(3\cos^2\theta - 1)(m-1/2)\nu_Q. \quad (17)$$

When the external field $H$ ($H \cdot H_N$) is applied alon the $c$-axis, then NMR frequency can be approximated to vary as $\nu_m + \gamma H\cos\beta/2\pi\hbar$, where $\beta$ is the angle between $\mathbf{H}$ and $\mathbf{H}_N$.

Using Eq. (16) we can obtain the following expression of the resonance field for the central NMR frequency $\nu_{-1/2\leftrightarrow1/2}$

$$H_{res} = H_F + (1/2)(3\cos^2\theta - 1)H_d' = H_N(\theta), \quad (18)$$

therefore we can analyze the resonance field $H_{res}$ by separating the dipolar contribution in accord with $\theta$ from the resonance frequency obtained.

## IV. INTERPRETATION OF $^{55}Mn$ NMR AND DISCUSSION
### A. $Mn^{4+}$ ion

The hyperfine field of the $Mn^{4+}$ ion deduced from the resonance frequency in Fig. 2 is 218.4 ±0.1 kG, which is appreciably smaller as compared with both the theoretically-calculated and the experimentally-observed values as we see below in other oxides. Theoretically the Fermi-contact hyperfine field per unpaired d-electron, which is defined by $\chi$, was calculated by Freeman and Watson, by summing up totally the polarized contribution of inner 1s, 2s and 3s electrons, for various valence states of free manganese ions.[47] The value of $\chi$ for $Mn^{4+}$ is −2.34 a.u.. Hence we find $H_F(calc)= 3\chi_{Mn4+}=$ −293 kG, using the conversion factor 1 a.u.=4.17×10 kG for $\chi$ in atomic unit. The observed hyperfine field for $Mn^{4+}$ ions in $Mn_{12}Ac$ is smaller by 26% than the calculated value. Moreover there is a similar evidence for the reduction of the hyperfine fields in $Mn^{4+}$ ions in $Mn_{12}$ cluster; the preliminary measurements on the $^{55}Mn$ NMR in $Mn^{4+}$ ion in isomorphous $[Mn_{12}O_{12}(O_2CPh)_{16}(H_2O)_4]\cdot 2PhCO_2H$ [54] yield almost same resonance frequencies as the present case. On the contrary, in lithium manganese ferrite $Li_{0.5}Mn_{0.5}Fe_2O_4$, the hyperfine fields for $^{55}Mn$ nuclei in $Mn^{4+}$ ions are observed to be −293.8±0.1 kG.[53] The hyperfine coupling constant $A$ concerning $Mn^{4+}$ ions in paramagnetic state [46] is reported to give $H_N$=−301.86 kG (for $Mn^{4+}$ ions diluted in MgO) and −290.90 kG (for $Mn^{4+}$ ions diluted in $Al_2O_3$). These experimental values for the hyperfine fields in $Mn^{4+}$ ions are nearly the same as the above theoretical one. So we can imply that $^{55}Mn$ NMR associated with the $Mn^{4+}$ ion in $Mn_{12}$ cluster is usually observed around 230 MHz ($H_N$=−218 kG) which corresponds to about 74% of the calculated value. The following fact should be noted; recent polarized-neutron-diffraction measurements on $Mn_{12}Ac$ also indicate that local magnetic moments for Mn1 sites are reduced to be 78±4 % of full value 3.0 $\mu_B$ for $Mn^{4+}$ ion and those for Mn2(Mn3) sites to be 92±4 %(94±3 %) of full value 4.0 $\mu_B$ for $Mn^{3+}$ ions.[34] The theoretical calculation on the electronic structure for $Mn_{12}Ac$ cluster based on density functional formalism has presented a considerable amount of reduction for the local magnetic moments of manganese ions; 2.6 $\mu_B$(86% of full value 3.0 $\mu_B$) for $Mn^{4+}$ ions and 3.6 $\mu_B$(90% of full value 4.0 $\mu_B$) for $Mn^{3+}$ ions, respectively.[14] The reduction factor of 26% for $Mn^{4+}$ ions observed in the present NMR measurement is consistent with that of the polarized neutron result. We believe a large amount of reduction of $Mn^{4+}$ hyperfine field found for $Mn_{12}Ac$ by present NMR measurement is related to the covalence and the strong exchange coupling between Mn1 and Mn2 via oxygen ions as mentioned in Ref. 14.

Next we interpret $^{55}Mn$ NMR spectra of the inset of Fig. 3. As is previously noted, the spectra $\nu_»$ arise from the $^{55}Mn$ nuclei in $Mn^{4+}$ ions belonging to the clusters whose moments are parallel to the external field while those of $\nu_«$ arise from ones belonging to the clusters whose moments are antiparallel to the external field. We find that the intensity of the spectra $\nu_»$ is always larger than that of $\nu_«$. This is due to the fact that the spin-echo decay time $T_2$ is longer for $\nu_»$ line than for $\nu_«$. As the time passes the spin-echo intensity of $\nu_»$ line has a trend to increase while that of $\nu_«$ line to decrease. This fact reflects the relaxation of cluster magnetization in the presence of the external field after ZFC; the increase of the intensity of $\nu_n$ line accompanying the decrease of that of $\nu_a$ line corresponds to the gradual change of the magnetization direction of the cluster from down to up with respect to the field direction during the time elapses. The fact that the slope of $\nu$-$H$ relation coincides with the value of $^{55}\gamma/2\pi$ means that the hyperfine field of the $Mn^{4+}$ ion is just parallel to the external field and its own local moment.

We mention here the small quadrupole-splitting $\Delta\nu_Q$=0.72±0.05 MHz. From the local symmetry of Mn1 site in $Mn_{12}Ac$ molecule [7] each $Mn^{4+}$ ion is subject to an octahedral crystal field with an axial-symmetry axis which is parallel to the $c$-axis. The electric-field gradient $V_{ZZ}$ at $^{55}Mn$ nuclei in $Mn^{4+}$ ions originate solely from octahedrally-surrounding $O^{2-}$ ions with symmetry lower than cubic one. [7] There may be second-order contribution to the quadrupole coupling that is due to the distortion of spherical electron shells of $Mn^{4+}$ ions induced by the external charges, and the induced gradient is given to be $-\gamma_\infty V_{ZZ}$ in which $\gamma_\infty$ is an anti-shielding factor. Then the total electric-field gradient is to be $q_c=(1-\gamma_\infty)V_{ZZ}$. From the value of $\Delta\nu_Q$= $3e^2q_cQ/\{2I(2I-1)h\}$ =0.72±0.05 MHz we find $e^2q_cQ/h$=4.80±0.33 MHz.

### B. $Mn^{3+}$ ion

Since the orbital doublet ion such as $Mn^{3+}$ ion stabilizes in energy by lowering the symmetry of ligands to remove the degeneracy in the cubic symmetry, the ground-state wave function becomes asymmetric, which arises considerable amount of the dipolar hyperfine field on a nuclear site. In fact as for the $Mn^{3+}$ ions both at Mn2 and at Mn3 sites, the octahedral ligands composed of oxygens are of tetragonal symmetry in the first approximation and in addition have lower symmetry than tetragonal one. As shown in Fig. 1(b), the $Mn^{3+}$ ion at the Mn2 site has mainly tetragonal symmetry (elongated) with the axially symmetric axis $\delta$, which lies in the direction of 11.7° from the $c$-axis. On the other hand, the $Mn^{3+}$ ion at the Mn3 site has considerable rhombic component [7] in addition to the tetragonal one, in which the $\delta$-axis makes an angle 36.2° from the $c$-axis.

As is already noted, the hyperfine field of the $Mn^{3+}$ ion has an anisotropic dipolar term $H_d$ besides an isotropic Fermi-contact term $H_F$. As is clear from Eq. (16) the $^{55}Mn$ NMR frequency for the $Mn^{3+}$ ion varies appreciably depending on two frequencies $\nu_F$ and $\nu_d$ and an angle $\theta$ shown in Fig. 1(b) and Fig. 4(c). We next evaluate the two parameters, $H_F$ and $H_d$, of the hyperfine field. The value of $\chi$ calculated for free $Mn^{3+}$ ion is given to be −2.91 a.u. [47] which corresponds to $4\chi_{Mn3+}$ =−485 kG≡$H_F$(calc) for the calculated value of Fermi-contact term in the $Mn^{3+}$ ion. On the other hand the dipolar hyperfine field parameter in Eq. (A4) is given to be $h_d = (4/7)\mu_B\langle r^{-3}\rangle = (2/7)\times 125\langle r^{-3}\rangle_{a.u}$ [46]$=+171$ kG $\equiv h_d$(calc), using $\langle r^{-3}\rangle=4.79$ a.u. for free $Mn^{3+}$ ion.[47]

However it turns out that the use of these theoretical values fails to explain our NMR results. Then in order to obtain to find reasonable agreement between the experiment and calculation it is necessary to take into account the reduction factor for the Fermi-contact and the dipolar hyperfine fields. Here as for the dipolar hyperfine field we apply the reduction factor for 3d-electron spins obtained by neutron diffraction measurement because the dipolar hyperfine field is directly related to the local moment of the ion. Using the experimental value of the $Mn^{3+}$ magnetic moment 3.69±0.04 $\mu_B$ (8±1 % reduced from 4.0 $\mu_B$) at Mn2 which was obtained by the polarized neutron measurement, [34] we can estimate the dipolar hyperfine field parameter to be $h_d = h_d$(calc)×0.92 $= +157.3$ kG for the $Mn^{3+}$ ion at Mn2. Next we take $H'_d = h_d$ by assuming the tetragonal symmetry of six-fold $O^{2-}$ ligands around the $Mn^{3+}$ ion at Mn2 without any other lower symmetry. In order to obtain the experimental value $H_N=-264.8$ kG from Eq. (18) by using the following values, $H'_d=+157.3$ kG and $\theta=11.7°$, it is necessary to choose the Fermi-contact hyperfine field to be $H_F=-412.4$ kG for $Mn^{3+}$ ion at Mn2. This value corresponds to 85% of the above-calculated value 485 kOe. These values of hyperfine fields, $H_F=-412.4$ kG and $H'_d=+157.3$ kG, are comparable to $H_F=-414.8$ kG and $H'_d=+113.6$ kG deduced from ESR measurement for the $Mn^{3+}$ ion in $TiO_2$.[51]

Since it is clear in general that $H_F$ is roughly constant for ions of common valence in common environments, [47,56] we here reasonably adopt the same value of $H_F$ to the $Mn^{3+}$ ion at the Mn3 site. Then by substituting the observed value of $H_N=-345.4$ kG, $H_F=-412.4$ kG and $\theta=36.2°$ in Eq. (18) we find $H'_d=+140.5$ kG for the $Mn^{3+}$ ion at the Mn3 site. As is shown in Eq. (A4) the effective dipolar hyperfine field parameter $H'_d=h_d\cos 2\phi$ in dipolar hyperfine coupling tensor D varies depending on the ratio of admixing of $|3Z^2-r^2\rangle$ wave function to $|X^2-Y^2\rangle$ as specified by Eq. (12). The obtained result $H'_d=+140.5$ kG for Mn3 can be reasonably interpreted taking into account admixing coefficients for the ground-state wave function, $\cos\phi$ and $\sin\phi$, to satisfy the relation $\cos 2\phi = 0.893$ ($\phi=13.4°$). The quantitative interpretation of hyperfine fields for the $Mn^{3+}$ ions in $Mn_{12}Ac$ is arranged in Table I. For the case of the $Mn^{3+}$ ion at the Mn3 the ground-state wave function has to be defined under the influence of considerable rhombic component of the crystal field, which results in the smaller dipolar hyperfine field in terms of the admixing coefficient $\cos 2\phi$. The quadruple coupling constant $e^2q_vQ/h$ can be deduced from $\breve{\nu}_Q= 4.3\pm0.1$ for $^{55}Mn$ at Mn2 and $2.9\pm0.1$ MHz for $^{55}Mn$ at Mn3 to be $30.6\pm0.7$ MHz and $40.6\pm1.5$ MHz, respectively.

We have analyzed the hyperfine fields of $Mn^{4+}$ and $Mn^{3+}$ ions in $Mn_{12}Ac$ and quantitatively clarified that the former only has an isotropic term $H_F$ while the latter has an anisotropic terms $H_d$ in addition to $H_F$. Using these parameters we can deduce total angular dependence of the hyperfine fields for $Mn^{3+}$ ions as expressed in Eq. (13). The obtained results from the present NMR studies are summarized in Table II. For the case of $Mn^{3+}$ ions (both Mn2 and Mn3) the total hyperfine fields $\mathbf{H}_N$ are not parallel to their own spins except for $\breve{\theta}=0$ and $\pi/2$, as noted in Eq. (14), but make angle $\alpha - \breve{\theta}$ from the $c$-axis, which causes $^{55}$Mn NMR frequency $\nu_{-1/2 \leftrightarrow 1/2} = {}^{55}\gamma |\mathbf{H}_N+\mathbf{H}|/2\pi$ in the external field approximately to be $\{^{55}\gamma H_N \pm {}^{55}\gamma \cos(\alpha - \breve{\theta})H\}/2\pi$ in the condition of $H \ll H_N$. Therefore the effective value of $\gamma/2\pi$, $\gamma_{eff}/2\pi$, obtained from the frequency-field diagrams for $Mn^{3+}$ ion in Fig. 3 is expected to be $(^{55}\gamma/2\pi)\cos(\alpha - \breve{\theta})$. The fact that the value $\cos(\alpha - \breve{\theta})$ evaluated from the values of $\breve{\theta}$ and $\alpha$ in Table II deviates less than 0.04 from 1 implies the value of $\gamma_{eff}/2\pi$ for $^{55}$Mn NMR in $Mn^{3+}$ ions would be quite similar to that of $^{55}\gamma/2\pi$ of free $^{55}$Mn nucleus. This is the case just observed in Fig. 3 for $^{55}$Mn NMR in $Mn^{3+}$ ions.

Quite recently Furukawa *et al*[45] presented a report on the zero-field $^{55}$Mn NMR and field-dependence of the NMR frequencies for oriented powder $Mn_{12}Ac$, together with nuclear relaxation rate $1/T_1$. They discussed nuclear relaxation by thermal fluctuations in the $S=10$ spin states, and from the frequency-field diagram they denoted the standard picture of magnetic structure in the cluster. They assumed that the dominant part of the hyperfine field of the $^{55}$Mn nuclei in $Mn^{3+}$ ion was core-polarization term, neglecting dipolar terms. Accordingly the identification of two spectra concerned with Mn2 and Mn3 remains unsolved. Here we have reasonably identified these two spectra around 280 MHz and 364 MHz, by taking into account the contribution from the dipolar hyperfine field, to be due to $Mn^{3+}$ ions at Mn2 and Mn3, respectively.

### C. Hyperfine interaction in $Mn_{12}Ac$ cluster

We discuss here the total hyperfine interaction between manganese nuclei and the molecular spin $S=10$ in $Mn_{12}Ac$ by taking into account three kinds of hyperfine coupling constants obtained for Mn1, Mn2 and Mn3, respectively. The roles of the hyperfine interaction in the magnetic relaxation of $Mn_{12}Ac$ clusters have been discussed by several authors.[10–13,16] The hyperfine interactions concerned with eight $Mn^{3+}$ and four $Mn^{4+}$ ions in $Mn_{12}Ac$ cluster has been studied quantitatively by Hartmann-Boutron *et al*[2] assuming two kinds of hyperfine coupling constants for manganese ions. Since the present $^{55}$Mn NMR results in $Mn_{12}Ac$ show us three kinds of ground-states for single manganese ions, we express the total magnetic hyperfine interaction in the $Mn_{12}$ cluster in terms of three kind of hyperfine tensors [46] $\mathbf{A}_{xyz}$ for single ions as

$$\langle H_n^{tot}\rangle = \sum_{r=1}^{12}\langle H_n\rangle_r = \sum_{i=1}^{4}\mathbf{S}_i^{(1)}\mathbf{A}_i^{(1)}\mathbf{I}_i^{(1)} + \sum_{j=1}^{4}\mathbf{S}_j^{(2)}\mathbf{A}_j^{(2)}\mathbf{I}_j^{(2)} + \sum_{k=1}^{4}\mathbf{S}_k^{(3)}\mathbf{A}_k^{(3)}\mathbf{I}_k^{(3)} \quad (19)$$

where superscripts (1), (2) and (3) concern with Mn1, Mn2 and Mn3 sites, respectively. Each sum is limited over manganese ions within one of Mn1, Mn2 and Mn3 sites. Next we take into account the ferrimagnetic structure in the $Mn_{12}$ cluster where the quantization axis is taken to be as $z$-axis along the $c$-axis, and assuming in the ground-state the $x$ an $y$ component of the electron spin is replaced by 0. Then Eq. (19) can be simplified to be

$$\langle H_n^{tot}\rangle_{ferri} = S_z^{(1)}\mathbf{A}_{zz}^{(1)}\sum_{i=1}^{4}I_{zi}^{(1)} + S_z^{(2)}\mathbf{A}_{zz}^{(2)}\sum_{j=1}^{4}I_{zj}^{(2)} + S_z^{(3)}\mathbf{A}_{zz}^{(3)}\sum_{k=1}^{4}I_{zk}^{(3)}$$

$$= +\tfrac{3}{2}\mathbf{A}_{zz}^{(1)}\sum_{i=1}^{4}I_{zi}^{(1)} - 2\mathbf{A}_{zz}^{(2)}\sum_{j=1}^{4}I_{zj}^{(2)} - 2\mathbf{A}_{zz}^{(3)}\sum_{k=1}^{4}I_{zk}^{(3)}. \quad (20)$$

By use of the relation $\nu = A_{zz}S/h$ from the observed central frequencies $\nu^{(1)}=230$, $\nu^{(2)}=280$ and $\nu^{(3)}=365$ MHz, we can deduce the hyperfine coupling constants divided by $h$, referring to $\mathbf{A}_{zz}^{(1)}/h$, $\mathbf{A}_{zz}^{(2)}/h$ and $\mathbf{A}_{zz}^{(3)}/h$ as $\mathbf{A}_z^{(1)}$, $\mathbf{A}_z^{(2)}$ and $\mathbf{A}_z^{(3)}$ to be 153.5, 139.7 and 182.2 MHz, respectively. Accordingly for the total hyperfine interaction at the ferrimagnetic ground-state of molecular $Mn_{12}Ac$, we obtain the following equation in MHz,

$$(\frac{1}{h})\langle H_n^{tot}\rangle_{ferri} = +\tfrac{3}{2}(153.5)\sum_{i=1}^{4}I_{zi}^{(1)} - 2(139.7)\sum_{j=1}^{4}I_{zj}^{(2)} - 2(182.2)\sum_{k=1}^{4}I_{zk}^{(3)}. \quad (21)$$

Every sum takes integer values from $-10$ to 10 considering $I(^{55}Mn)=\tfrac{5}{2}$. The number of allowed value of the energy is $(2\times10+1)^3=9261$.

The maximum value of the effective hyperfine interaction for ferrimagnetic coupling in $Mn_{12}Ac$ molecule can be obtained from the next relation,

$$(\frac{1}{h})\langle H_n^{tot}\rangle_{ferri} = 2\cdot 10(\tfrac{3}{4}\times 153.3 + 139.7 + 182.2)\ \text{MHz} = 0.29\ \text{cm}^{-1}. \quad (22)$$

From the definition $\langle H_n^{tot}\rangle_{ferri} = g_z\mu_B S_z\cdot h_e$ we can estimate $h_e=324$ G, with using $g_z=1.93$, [35] for the maximum nuclear hyperfine field that is seen by the total spin $S=10$ in $Mn_{12}Ac$. This value amounts to 2.5 times as large as the nuclear hyperfine field for the $Mn^{4+}$ ion (137 G) in $Mn_{12}Ac$. Therefore every electronic energy-level specified by $|S,m\rangle$ is inhomogeneously broadened, through the hyperfine interaction with $^{55}Mn$ nuclei, into $21^3$ sub-levels with the total spread of about 0.3 cm$^{-1}$ (0.42 K). The broadened energy-levels which approximately appear like a continuum symbolically depicted in Ref. 2.

This amount of hyperfine interaction was also confirmed by measurements of magnetic specific heat at low temperatures where appreciable nuclear contribution appeared below 0.5 K in $Mn_{12}Ac$.[28,30]

Next we refer to the other components of the hyperfine coupling tensor $\mathbf{A}_{xyz}^{(p)}/h$ in MHz which play an important role in magnetization relaxation and also in $^{55}Mn$ nuclear relaxation in $Mn_{12}Ac$. For Mn1 the hyperfine interaction is isotropic, then $\mathbf{A}_{xyz}^{(1)}/h$ is 153 MHz. The hyperfine coupling tensor $A_{xyz}^{(2)}/h$ for Mn2 and $\mathbf{A}_{xyz}^{(3)}/h$ for Mn3 are given in Appendix B. When we put these values into Eq. (19) we can represent the total hyperfine interaction in the cluster. This type of the total hyperfine interaction in $Mn_{12}$ cluster is understood to be strictly anisotropic in character.

The above transverse component of the hyperfine interaction would contribute in the magnetic relaxation [2,12] of the cluster to promoting the transition between $|S,m\rangle$ levels as well as molecular-dipole interaction and higher-order crystal anisotropy. Luis, Baltolome and Fernandez [13] calculated the relaxation rate as a function of the longitudinal magnetic field taking into account dipolar fields among clusters together with hyperfine fields and the quartic magnetic anisotropy, and explained non-linear spin relaxation rate in terms of spin-phonon-induced tunneling depending on the longitudinal external field. On the other hand Leuenberger and Loss [15] neglecting the hyperfine and dipolar fields for $T>1$ K, calculated the relaxation rate as a function of the longitudinal magnetic field based on phonon-assisted spin tunneling induced by the quartic magnetic anisotropy and weak transverse magnetic fields. Their result was also in good agreement with experimental relaxation rates including all resonance peaks. It has been assured from these theoretical calculations that any extent of a transverse component is essential to explain the magnetic relaxation in $Mn_{12}$ cluster, no matter what the phonon-assisted-quantum transition may be induced by an external or internal origin.

### D. Relaxation of $Mn_{12}Ac$ cluster magnetization

As shown in Fig.5, our $^{55}Mn$ NMR result has demonstrated that the magnetization recovery follows square-root time dependence. In general the recovery curve of the cluster magnetization $M$ is expressed by a stretched-exponential function $M = M_0[1-\exp\{-(t/\tau)^\beta\}]$, where it has been shown theoretically that $\beta =1/2$ holds in the short-time regime and $\beta =1$ in the long-time regime, respectively.[10,11] The evidence for short-time relaxation has been observed experimentally for $Fe_8$ [56,57] at temperatures below 1 K in low fields and also for $Mn_{12}Ac$ in zero field [22,58] or with fields [44] below 2 K. It is noted that in our case the square-root decay of magnetization has been observed in nearly whole range of external magnetic fields between 0.200 T and 1.900 T at T=2.0 K. In the inset of Fig. 5 the gradual decrease of $\tau$ versus $H$ can be interpreted as the relaxation process governed by the spin-phonon interaction.[2,8] The occurrence of the regular minimums of $\tau$, at every 0.45 T which is nearly equal to $D/g\mu_B \equiv H_1$ using $D \approx 0.58$ K, implies that the resonant quantum tunneling [19,20] occurs between the levels of the two-wells having the same energy. This feature is found obviously above 2.0 K while below 1.8 K it is difficult to observe such a marked behavior. Taking into account such a thermal effect that the relaxation times depend strongly on temperatures, we understood that the appearance of dips correspond to phonon-assisted resonant quantum tunneling which has been discussed by several authors.[18,19] The relaxation time of the magnetization in Mn$_{12}$Ac, investigated in long-times by use of proton NMR by Jang *et al*,[38] has been measured as a function of applied magnetic field at T=2.4 K. Their result was interpreted as the background relaxation of exponential law due to thermal excitations over the barrier accompanying relaxation minimums at field values corresponding to magnetic level crossings. In our case the result of the relaxation time shown in inset to Fig. 5 indicates that the background relaxation, showing more abrupt decrease with the external field, does not follow strictly to the Arrhenius law. Such a deviation is due to the transeverse field effect caused by the small extent of inclination of the applied field from the easy axis which are occurred in the present oriented crystal.

Since there exists the forth-order term in Eq. (1), $-ES_z^4$, the value of the magnetic field at magnetic level crossing (resonant fields of quantum tunneling) is not the same for every level of $|S,m\rangle$ spin state.[15,16,30] Namely a level of $|S,m\rangle$ in the metastable potential well comes into resonance with the level of $|S,m'\rangle$ in the stable well at a longitudinal magnetic field $H_Z = nH_1\left[1 + \frac{E}{D}\left(m^2 + m'^2\right)\right]$ strictly not at $H_Z=nH_1$ ($H_1=D/g\mu_B$). In other words, this means that there appear some satellites for the maximum of the relaxation rate, depending on $m$ and $m'$ in the vicinity of the main resonance fields $H_Z=nH_1$. It has been pointed out also that the satellite maximums become visible if the high resolution is achieved.[15,16] Recent measurements have succeeded in showing these satellites by the way of decreasing temperature[24,26] down to around 0.5 K or point-to-point precise measurements on relaxation rates with increasing temperature.[25] Though the present result in the inset of Fig. 5 is not sufficient in resolution to detect the satellites we try to observe them in future.

## V. CONCLUSION

By an analysis of the hyperfine fields of $^{55}$Mn nuclei associated with $Mn^{3+}$ and $Mn^{4+}$ ions, we have interpreted three kinds of $^{55}$Mn NMR composed of five-fold quadrupole-split lines and identified them to the corresponding sites in $Mn_{12}Ac$ cluster. Namely the central frequencies 230.2±0.1 MHz, 279.4±0.1 MHz and 364.4±0.1 MHz were assigned to be $^{55}$Mn NMR originating in $Mn^{4+}$ ions at Mn1 sites, $Mn^{3+}$ ions at Mn2 sites and $Mn^{3+}$ ions at Mn3 sites in $Mn_{12}Ac$ molecule, respectively.

In $Mn_{12}Ac$ crystal the hyperfine field concerned with the $Mn^{4+}$ ion has been found to be predominantly reduced (74%) compared with the ones in other compounds. The hyperfine field of the $Mn^{4+}$ ion is isotropic in character because of being mainly due to Fermi-contact term while that of $Mn^{3+}$ is anisotropic and demonstrates contribution of the predominant dipolar term to the total hyperfine field which is deduced from the admixed wave function relevant to the low symmetry of the crystal field. The magnitude of the total hyperfine interaction in the $Mn_{12}$ cluster at the ferrimagnetic ground-state is estimated to be 0.3 cm$^{-1}$ at most using twelve of hyperfine constants of single manganese ion which are deduced from present NMR results. Therefore the exact energy-levels which are specified by $|S,m\rangle$ and determined by the spin Hamiltonian (1), are broadened like a continuum with a half width of 0.3 cm$^{-1}$ owing to the hyperfine interaction. The amount of above hyperfine interaction corresponds to the nuclear hyperfine field $h_e$=0.32 kG acting on $Mn_{12}$ cluster spin. The total hyperfine coupling tensor for Mn12 cluster has been obtained by summing up twelve of hyperfine coupling tensors of single ions that are composed of both isotropic ($Mn^{4+}$) and anisotropic ($Mn^{3+}$) components.

As for magnetic relaxation in $Mn_{12}Ac$ the cluster magnetization relaxes non-exponentially in short-times, i.e. it shows $\sqrt{t}$-recovery curve, at 2.0 K. The fact that the longitudinal-field dependence of the relaxation time in short-times shows significant dips of more than one order of magnitude at every 0.45 T step at T=2.0 K demonstrates that the thermally-assisted resonant quantum tunneling is active in early stage of magnetic relaxation at T=2.0 K. By using $^{55}$Mn NMR the magnetization relaxation concerning $Mn_{12}$ cluster in the fields can be measured very shortly after the reversal of the applied field. The measurements on the relaxation rates which increase significantly with application of measurable transverse applied fields are now progressing.

. **ACKNOWLEDGEMENT**

We thank M. Chiba for fruitful discussions. We thank also S. Miyashita for continual interests and comments. Thanks are due to B. Imanari for assistances in final stage of this work. Finally we thank F. Borsa and Y. Furukawa for showing a manuscript prior to publication.

## APPENDIX A: DIPOLAR HYPERFINE FIELD

In order to calculate the dipolar hyperfine field from Eq. (4) for the $Mn^{3+}$ ion by using the ground-state wave function (12), we should transform dipolar hyperfine Hamiltonian (4) by adopting the operator equivalent method to an electron configuration of a hole plus a half-filled shell which satisfies the *LS*-coupling scheme. Then the dipolar hyperfine Hamiltonian is expressed as follows [59]

$$H_d = \frac{2\gamma\hbar\mu_B}{r^3}\xi\left\{L(L+1)\mathbf{S}\cdot\mathbf{I} - \frac{3}{2}(\mathbf{L}\cdot\mathbf{S})(\mathbf{L}\cdot\mathbf{I}) - \frac{3}{2}(\mathbf{L}\cdot\mathbf{I})(\mathbf{L}\cdot\mathbf{S})\right\} \quad (A1)$$

where $\xi = \frac{2l+1-4S}{S(2l-1)(2l+3)(2L-1)}$.

Considering that $\psi_g$ contains $|2\rangle$, $|0\rangle$ and $|-2\rangle$ in terms of $|L_z\rangle$ notation, the dipolar hyperfine operator can be calculated in the following form by use of *XYZ*-frame notation, remaining only diagonal terms,

$$H_d = -\frac{2\gamma\hbar\mu_B}{r^3}\xi[\{L(L+1) - 3L_Z^2\}(S_Z I_Z - \tfrac{1}{2}S_X I_X - \tfrac{1}{2}S_Y I_Y) - \tfrac{3}{2}iL_Z(S_X I_Y - S_Y I_X)]. \quad (A2)$$

Then the effective dipolar hyperfine interaction can be obtained to be, using $\xi = \frac{-1}{42}$ for $S=2$, $l=2$ and $L=2$,

$$\langle\psi_g|H_d|\psi_g\rangle = \cos 2\phi \frac{2\gamma\hbar\mu_B}{7}\langle r^{-3}\rangle(S_Z I_Z - \tfrac{1}{2}S_X I_X - \tfrac{1}{2}S_Y I_Y). \quad (A3)$$

By equating Eq. (A3) to the expression $-\gamma\hbar\mathbf{I}\cdot\mathbf{H}_d$, we get the dipolar hyperfine field for the present case to be, in terms of a unit vector $\hat{\mathbf{e}}_m$ of magnetization $\mathbf{m}$ of $Mn^{3+}$ ion,[49]

$$\mathbf{H}_d = \mathbf{D}\,\hat{\mathbf{e}}_m \quad (A4)$$

where $\mathbf{D} = H'_d \begin{pmatrix} \frac{-1}{2} & 0 & 0 \\ 0 & \frac{-1}{2} & 0 \\ 0 & 0 & 1 \end{pmatrix}$ in terms of an effective dipolar hyperfine field parameter

$H'_d = h_d \cos 2\phi$ with a parameter $h_d = \frac{4\mu_B}{7}\langle r^{-3}\rangle$.

## APPENDIX B: HYPERFINE COUPLING TENSOR FOR $Mn^{3+}$ ION

The hyperfine interaction for the Mn$^{3+}$ ion can be expressed in the frame of local tetragonal axes (principal axes) *XYZ* as follows,

$$\langle \psi_g | H_{\text{hyp}} | \psi_g \rangle = \mathbf{S} \cdot \mathbf{A}_{XYZ} \cdot \mathbf{I} \tag{B1}$$

with

$$\mathbf{A}_{XYZ} = \tfrac{1}{2} \gamma \hbar |H_F| \mathbf{1} - \tfrac{1}{2} \gamma \hbar \mathbf{D}, \tag{B2}$$

where $\mathbf{1}$ is a unit tensor. This relation is analogous to Eq. (13).

However, since in Mn$_{12}$Ac cluster the quantization axes of spins are defined by the crystal *c*-axis we have to express the above hyperfine coupling tensor in *xyz*-frame where *z*-axis is taken as the *c*-axis. The result of transformed notation of the hyperfine coupling tensor from *XYZ*-frame to *xyz*-frame shown in Fig. 4(b) is as follows,

$$\langle \psi_g | H_{\text{hyp}} | \psi_g \rangle = \mathbf{S} \cdot \mathbf{A}_{xyz} \cdot \mathbf{I} \tag{B3}$$

with

$$\mathbf{A}_{xyz} = \tfrac{1}{2} \gamma \hbar |H_F| \mathbf{1} - \tfrac{1}{2} \gamma \hbar \mathbf{D}', \tag{B4}$$

where

$$\mathbf{D}' = H_d' \begin{pmatrix} \frac{2-3\cos^2\theta}{2} & 0 & \frac{-3}{4}\sin 2\theta \\ 0 & -\frac{1}{2} & 0 \\ \frac{-3}{4}\sin 2\theta & 0 & \frac{3\cos^2\theta-1}{2} \end{pmatrix}. \tag{B5}$$

Accordingly we get

$$\mathbf{A}_{xyz} == \tfrac{1}{2} \gamma \hbar |H_F| \begin{pmatrix} 1 - \frac{2-3\cos^2\theta}{2} k & 0 & \frac{-3}{4}\sin 2\theta \cdot k \\ 0 & 1 & 0 \\ \frac{-3}{4}\sin 2\theta \cdot k & 0 & 1 - \frac{3\cos^2\theta-1}{2} k \end{pmatrix} \tag{B6}$$

with $k = H_d' / |H_F|$.

Then with using suitable values of $\theta$, $H_F$ and $H_d$ in TABLE 2, the hyperfine coupling tensors $\mathbf{A}_{xyz}^{(2)}$ for Mn2 and $\mathbf{A}_{xyz}^{(3)}$ for Mn3 are obtained in MHz as follows,

$$\mathbf{A}_{xyz}^{(2)} = \begin{pmatrix} 253.9 & 0 & -24.7 \\ 0 & 176.1 & 0 \\ -24.7 & 0 & 139.7 \end{pmatrix} \tag{B7}$$

and

$$\mathbf{A}_{xyz}^{(3)} = \begin{pmatrix} 201.9 & 0 & -53.0 \\ 0 & 180.5 & 0 \\ -53.0 & 0 & 182.2 \end{pmatrix}, \text{respectively.} \tag{B8}$$

**Footnote**

*Corresponding author. Electronic address: kubotk@nara-edu.ac.jp